\documentclass[pre,aps,amsmath,amssymb,showpacs]{revtex4}
\usepackage{graphicx,epsfig}
\usepackage{wasysym}

\begin{document}

\title{Thermal Degradation of Unstrained Single Polymer Chain:\\ Non-linear
Effects at Work}

\author{J. Paturej$^{1,2}$, A. Milchev$^{1,3}$,  V. G. Rostiashvili$^1$, and T.A.
Vilgis$^1$}
\affiliation{$^1$ Max Planck Institute for Polymer Research, 10 Ackermannweg,
55128 Mainz, Germany\\
$^2$ Institute of Physics, University of Szczecin, Wielkopolska 15,
70451 Szczecin, Poland
\\$^3$ Institute for Physical Chemistry, Bulgarian Academy of Sciences, 1113
Sofia, Bulgaria}

\begin{abstract}
We examine the thermally-induced fracture of an unstrained polymer chain of
discrete segments coupled by an anharmonic potential by means of Molecular
Dynamics simulation with a Langevin thermostat. Cases of both  under- and
over-damped dynamics are investigated, and a comparison with recent studies of
bond scission in model polymers with harmonic interactions is performed. We
find that the polymer degradation changes qualitatively between the inertial
regime and that of heavily damped dynamics. The role of bond healing
(recombination) is also studied and probability distributions for the
recombination times and overstretched bond lengths are obtained.

Our extensive simulations reveal many properties of the scission dynamics in
agreement with the notion of random breakdown of independent bonds, e.g., the
mean time of chain rupture, $\langle \tau \rangle$ follows an Arrhenian behavior
with  temperature $T$, and depends on the number of bonds $N$ in the polymer as
$\langle \tau \rangle \propto N^{-1}$. In contrast, the rupture rates of the
individual bonds along the polymer backbone indicate clearly the presence of
self-induced inhomogeneity  resulting from the interplay of  thermal noise
and nonlinearity.

Eventually we examine the fragmentation kinetics during thermolysis. We
demonstrate that both the probability distribution function of fragment sizes
as well as the mean length of fragments at subsequent times $t$ characterize
degradation as predominantly a first order reaction.
\end{abstract}

\maketitle

\section{Introduction}

The study of degradation and stabilization of polymers is important both from
practical and theoretical viewpoints \cite{Allen}. Disposal of plastic wastes
has grown rapidly to a world problem so that increasing environmental concerns
have prompted researchers to investigate plastics recycling by degradation as
an alternative \cite{Madras}. On the other hand,  degradation of polymers in
different environment is a major limiting factor in their application. Thermal
degradation (or, thermolysis) plays a decisive role in the design of
flame-resistant polyethylene and other plastic materials \cite{Nyden}. It
can also be used in conjunction with chromatography to characterize polymeric
structure \cite{Flynn}. Recently, with the advent of exploiting biopolymers as
functional materials \cite{Schulten,Han} the stability of  such materials has
become an issue of primary concern.

Most theoretical investigations of polymer degradation have focused on
determining the rate of change of average molecular weight
\cite{Simha,Jellinek,Ballauff,Ziff,Cheng,Nyden2,Doerr,Wang,Hathorn,Doruker}.
The main assumptions of the theory are that each link in a long chain molecule
has equal strength and equal accessibility, that they are broken at random, and
that the probability  of rupture is proportional to the number of links present.
Experimental study of polystyrene, however, have revealed discrepancies
\cite{Jellinek} with the theory \cite{Simha} so, for example, the thermal
degradation stops completely or slows down markedly when a certain chain length
is reached. Thus, all of the afore-mentioned studies investigate exclusively the
way in which the distribution of bond rupture probability along the polymer
backbone affects the fragmentation kinetics and the distribution of fragment
sizes as time elapses. Only few theoretic studies \cite{Lee,Sokolov} have
recently explored how does the single polymer chain's dynamics itself affect the
resulting bond rupture probability. In both studies, however, one has worked
with a phantom Gaussian chain (that is, one has used harmonic bond potentials
in the simulations) in order to linearize the problem and make it tractable by
some analytic approach like the multidimensional Kramers theory, used by Lee
\cite{Lee}, and the Wilemski-Fixman  approach, employed by Fugmann and
Sokolov \cite{Sokolov}. In addition, these investigations have been carried out
in the heavily damped regime of polymer dynamics where acceleration and
inertial effects are neglected.

In the present investigation we model the thermal degradation of a linear
polymer chain where monomers are connected by more realistic
non-linear (anharmonic) forces (Morse and Lennard-Jones)  using Langevin
molecular dynamics simulation in $3d$. By changing the friction coefficient of
the particles, $\gamma$, we examine the scission dynamics of the bonds in both
the under- and over-damped cases and find significant qualitative differences.

We investigate the average time of bond breakdown $\langle \tau \rangle$,
referred to frequently as the Mean First Breakage Time (MFBT) in the
literature, regarding its dependence on temperature $T$, on the number of bonds
$N$ in the polymer chain, or the friction $\gamma$ for both free chains as well
as for tethered chains with the one end fixed at a given position in space. We
produce also maps, presenting the  temporal evolution of various quantities
(strain, potential and kinetic energy, relative velocity) of each particular
segment or bond in an effort to detect indications for collective excitations
in the course of thermolysis. Also events of bond recombination (self-healing)
are investigated whereby the probability distribution of distances, traveled
by the affected monomers and times, elapsed between scission and recombination
are obtained. While many properties of the thermal degradation process are in
agreement with the notion of randomly breaking bonds, the obtained rate
histograms of bond rupture indicate unambiguously that the interplay of
noise and non-linear
interactions are responsible for a certain kind of self-induced
{\em multiple-scale-length inhomogeneity}
regarding the position of the breaking bonds along the backbone of the chain.

The paper is organized as follows: after this brief introduction, we sketch our
model in Section \ref{Model}, and present our main results in Section
\ref{MD_results}. We end this report by a short discussion and some concluding
remarks in Section \ref{Conclusions}.

\section{The model} \label{Model}

We consider a $3d$ coarse-grained model of a polymer chain which consists of $N$
repeatable units (monomers) connected by bonds, whereby each bond of length
$b$ is described by a Morse potential,
\begin{equation}\label{Morse_pot}
U_M(r) = D \{\exp[- 2 \alpha (r- b) - 2 \exp[-  \alpha (r- b)\}
\end{equation}
with a parameter $\alpha \equiv 1$. The dissociation energy of such bonds is
$D$, measured in units of $k_BT$, where $k_B$ denotes the Boltzmann constant and
$T$ is the temperature. The maximum restoring force of the Morse potential,
$f^{max} = -dU_M/dr = \alpha D / 2$, is reached at the inflection point, $r =
b+\alpha^{-1} \ln(2)$. This force $f^{max}$  determines the maximal tensile
strength of the chain. Since  $ U_M(0) = \exp(2\alpha b) - 2\exp(\alpha b)
\approx 1.95$,  the  Morse potential, Eq.~(\ref{Morse_pot}),  is only weakly
repulsive and beads could penetrate one another. Therefore, in order to allow
properly for the {\em excluded volume} interactions between bonded monomers, we
take the bond potential as a sum of $U_M(r)$ and  the so called
Weeks-Chandler-Anderson (WCA) potential, $U_{WCA}(r)$, (i.e., the shifted and
truncated repulsive branch of the Lennard-Jones potential);
\begin{equation}\label{WCA_pot}
U_{WCA}(r) = 4\epsilon \left [\left( \frac{\sigma}{r}\right )^{12} -
\left( \frac{\sigma}{r}\right )^{6} + \frac{1}{4}\right ] \Theta(2^{1/6}
\sigma - r))
\end{equation}
with $\Theta(x) = 0\; \mbox{or}\; 1$ for $x < 0$ or $x \ge 0$, and $\epsilon =
1$. The non-bonded interactions between monomers are also taken into account by
means of the WCA potential, Eq.~(\ref{WCA_pot}). Thus the interactions in our
model correspond to good solvent conditions. Thus, the length scale is set by
the parameter $\sigma = 1$ whereby the the monomer diameter $b \approx
2^{1/6}\sigma$.

In our MD simulation we use a Langevin equation, which describes the Brownian
motion of a set of interacting particles whereby the action of the solvent is
split into slowly evolving viscous force and a rapidly fluctuating  stochastic
force:
\begin{equation}\label{Langevin_eq}
 m \overrightarrow{\dot{v}}_i(t) = - \zeta \vec{v}_i + \vec{F}_M^i(t) +
\vec{F}_{WCA}^i(t) + \vec{R^i}(t) .
\end{equation}
The random force which represents the incessant collision of the monomers with
the solvent molecules satisfy the fluctuation-dissipation theorem $\langle
R^i_{\alpha}(t) R^j_{\beta}(t') \rangle = 2\zeta k_B T \delta_{ij}
\delta_{\alpha \beta}\delta(t-t')$. The friction coefficient $\zeta$ of the
Langevin thermostat, used for equilibration, has been set at $0.25$. The
integration step is $0.002$ time units (t.u.) and time is measured in units of
$\sqrt{m/\sigma^2 D}$ where $m$ denotes the mass of the beads.

We start the simulation with a well equilibrated conformation of the chain as a
random coil and examine the thermal scission of the bonds for both a free chain
as well as for a chain that is tethered with the end-monomer to a fixed position
in space. We measure the elapsed time $\tau$ until a bond rupture occurs, and
average these times over more than $2\times 10^4$ events so as to determine the
mean $\langle \tau \rangle$ which we also refer to as Mean First Breakage Time
(MFBT). In the course of the simulation we also sample the probability
distribution of breaking bonds regarding their position in the chain (a rupture
probability histogram), the probability distribution of the First Breakage Time,
$\tau$, the kinetic and potential energy of all bonds as well as the local
strain along the chain.

Since in the problem of thermal degradation there is no external force acting
on the chain ends, a well defined activation barrier for a bond scission is
actually missing, in contrast to the case of applied tensile force. Therefore, a
definition of an unambiguous criterion for  bond breakage is not self-evident.
Moreover, depending on the degree of stretching, bonds may break and then
recombine again. Therefore, in our numeric experiments we use a sufficiently
large expansion of the bond, $r_h = 5 b$, as a threshold to a broken state of
the bond. This convention is based on our checks that the probability for
recombination of bonds, stretched beyond $r_h$, is vanishingly small, i.e.,
below $1\%$.

\section{Simulation Results}
\label{MD_results}

\subsection{Dependence of the MFBT $\langle \tau \rangle$ on chain length $N$
and temperature $T$}

Our consideration of the $\langle \tau \rangle$ vs. $N$ dependence is based
on the assumption that bonds in the chain break entirely at random and the
scission events happen independent from one another \cite{Sokolov}. Consider
the survival probability  of the $i$-the bond in the chain, $S_i (t)$,  (i.e.,
the probability that after elapsed time $t$ the bond $i$ is still intact).
$S_i(t)$ may be written as  $S_i (t) = \exp (- \omega_i t)$, where $\omega_i$
is the corresponding scission rate of bond $i$. Then, for presumably random and
independent scission events, the survival probability of the total chain reads
\begin{equation}\label{Survival}
 S (N, t) = \prod_{i=1}^N S_i (t) = \exp \left( - N {\bar \omega} t\right)
\end{equation}
where the average bond scission rate is given by
\begin{equation}\label{Average}
{ \bar \omega} = \dfrac{1}{N} \sum_{i=1}^N  \; \omega_i
\end{equation}
Thus, the MFBT $\langle \tau \rangle$ of the whole chain can be represented as
\begin{equation}\label{MFBT}
\langle \tau \rangle = - \int\limits_{0}^{\infty} t \: \dfrac{\partial}{\partial t} \: S (N, t) =
\dfrac{1}{N {\bar \omega}}
\end{equation}
where we have used the general relationship between the survival probability and
the mean first passage time (see, e.g., Sec. 5.2.7 in ref. \cite{Gardiner}). It
is worth  noting  that the product of probabilities in Eq. (\ref{Survival})
corresponds to the  well known {\it mean field approximation} in the theory of
phase transitions where one neglects the correlations  \cite{Goldenfeld}.

\begin{figure}[ht]
\vspace{0.5cm}
\begin{center}
\includegraphics[scale=0.4]{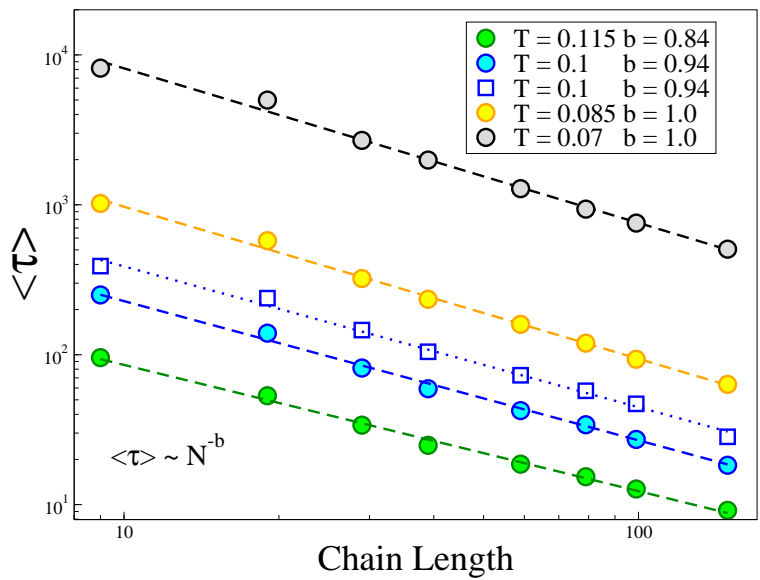}
\caption{Mean first passage time $\langle\tau\rangle$ vs chain length $N$ for
different temperatures of the heat bath ($\gamma = 0.25$). Dashed / dotted lines
with a slope $\approx -1.0$ represent fitting curves for grafted/free chains.
\label{tau_N_fig}}
\end{center}
\end{figure}

Our MD simulation results concerning  the dependence of MFBT $\langle \tau
\rangle$ on chain length $N$ are  shown in Fig.~\ref{tau_N_fig}. Evidently, one
observes a power-law decrease, $\langle \tau \rangle \propto N^{-\beta}$, with
$\beta \approx 1.0 \pm 0.15$ regardless of temperature.

This finding confirms the basic assumption that bonds break entirely at random
and the scission events happen independent from one another. One may readily
verify from the semi-log plot, Fig.~\ref{tau_N_fig}, that this $\langle \tau
\rangle$ vs $N$-relationship holds for both free and tethered single chains at
all temperatures that have been examined. We should like to note here that this
$\langle \tau \rangle$ vs $N$ relationship in thermolysis is very different from
the case of polymer breakdown under tension. When external force is applied,
$\langle \tau \rangle$ becomes nearly independent of chain length $N$ as the
tensile strength increases \cite{Ghosh}. In that case the chain thermal breakage
has been treated within the multidimensional Kramers-Langer escape theory. It
has been shown \cite{Ghosh} that the process of bond scission is governed by a
single {\it collective unstable mode} which peaks at an ``endangered'' bond and
decays exponentially towards both chain's ends. This explains why in this case
the MFBT is scarcely affected by the chain length $N$.

\begin{figure}[ht]
\vspace{0.5cm}
\begin{center}
\includegraphics[scale=0.4]{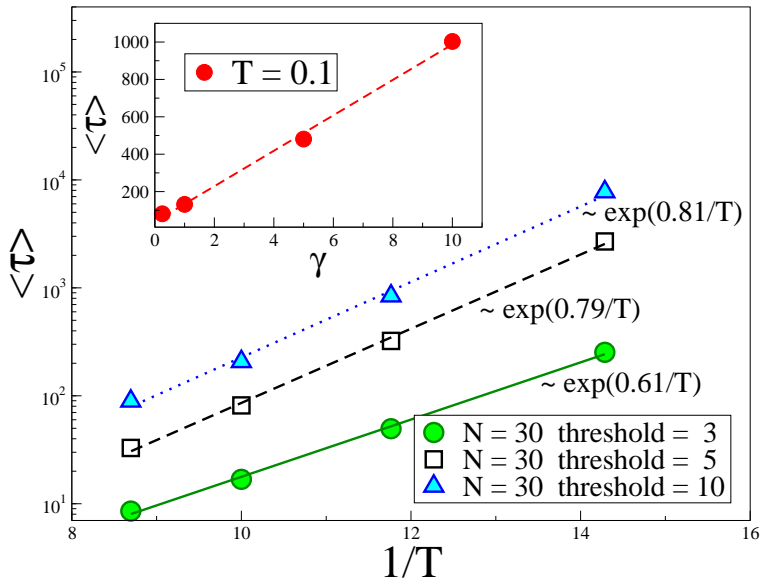}
\caption{Variation of the Mean First Breakage Time $\langle \tau \rangle$ with
inverse temperature $1/T$ for a chain with $N=30$ beads. Three different
thresholds for a broken bond  (given in the legend) have been examined. Fitting
lines correspond to an Arrhenian relationship, $\tau \propto \exp(\Delta E_b /
k_BT)$. The respective slopes $\Delta E_b$ are also indicated. The inset shows
the variation of $\langle\tau\rangle$ as a function of friction coefficient
$\gamma$ of the heat  bath. \label{Arrhenius_fig} }
\end{center}
\vspace{0.5cm}
\end{figure}

The dependence of the absolute value of the MFBT $\langle \tau \rangle$ on
inverse temperature, $1/T$, shown in  Fig.~\ref{Arrhenius_fig}, appears also  in
agreement with the general notion of polymer degradation as a thermally
activated process, 
Since, as mentioned in Section \ref{Model},  an unambiguous criterion for
scission is to a great extent arbitrary,  an appropriate  critical separation of
the beads that form a stretched bond, $r_h$, has to be chosen such that the
probability for recombination of a broken bond is negligible. With $r_h = 5$ we
find that less than $1\%$ of such bonds do subsequently recombine.

Fig.~\ref{Arrhenius_fig} demonstrates an expected variation of $\langle \tau
\rangle$  with (inverse) temperature, namely an
Arrhenian-dependence $\tau \propto
\exp(\Delta E_b / k_BT)$. Evidently, the measured ``activation barrier'' $\Delta
E_b$ depends weakly on the threshold for bond rupture $r_h$, and is found to
increase somewhat as  $r_h$ is varied in a broad interval $3 \le r_h \le 10$.
Note  that a threshold $r_h \approx 3$ corresponds to the position $r^{max}$ of
maximal tensile strength of the bond (i.e., the inflection point of the Morse
potential $U_M(r)$)  so that any value of $r_h \ge 3$ could be considered in
principle as appropriate for bond scission. Comparison of the results for
different rupture thresholds, Fig.~\ref{Arrhenius_fig}, indicates that for
$r_h \ge 5.0$ the energy barrier $\Delta E_b$ starts to be independent of $r_h$.
This allows us to estimate the activation barrier of thermally-induced scission
event to the value of around $0.8 D$ where $D$ is depth of the potential well,
Eq.~(\ref{Morse_pot}).

\subsection{Life time probability distribution}
\begin{figure}[ht]
\begin{center}
\includegraphics[scale=0.34,angle=270]{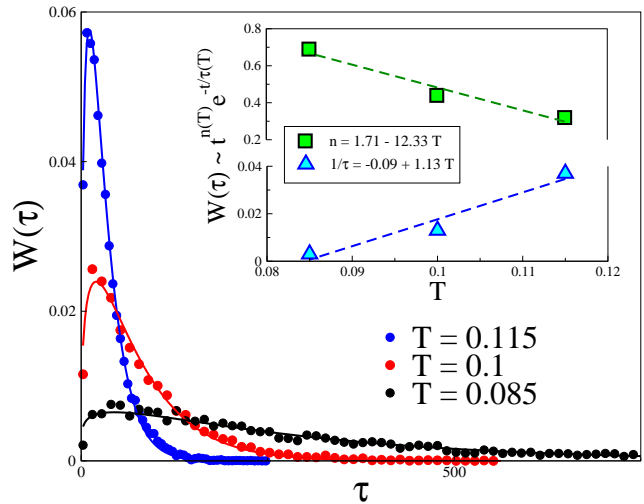}
\caption{First breakage time distributions $W(\tau)$ for a grafted chain with
$N=30$, $\gamma=0.25$ at different temperatures as indicated.
Symbols denote simulation result and full lines stand for the fitting function
$W(t) \propto t^n \exp (-t/\tau)$. The temperature dependence of $n$ and
$\tau^{-1}$ is shown in the inset. \label{PDF_tau_fig}}   \end{center}
\vspace{0.5cm}
\end{figure}

In Fig.~\ref{PDF_tau_fig} we present the probability distribution function
$W(\tau)$ of the random scission times $ \tau$ for several temperatures $T$
indicated in the legend.  Evidently, with decreasing temperature the rupture
times $\tau$ rapidly spread over a broad time interval even though the
location of the PDF maximum hardly changes.
It appears that $W(\tau)$ is well described by a Poisson - distribution function
$W(t) \propto t^{n(T)}\exp(- t/ \tau(T))$ (as one would
expect for uncorrelated random events)  with  temperature - dependent exponents
$n(T)$ and $\tau(T) $. In the inset to Fig.~\ref{PDF_tau_fig} one
may verify that both exponents change linearly with $T$. Evidently, the width of
$W(\tau)$, which  determines the variance of the life time $ \tau(T) $, declines
with growing temperature of the system.

\subsection{Self-healing bonds}
Concluding this first set of results, we consider briefly the ``self-healing''
of broken bonds during thermal degradation since it helps to determine the
appropriate criterion for a given bond to be declared broken. As in our previous
work on chain scission in the presence of constant load \cite{Ghosh}, we studied
the process of recombination (self-healing) after a given bond has exceeded the
threshold length $h_r$. In thermolysis the effect of self-healing should be even
more pronounced due the the absence of external force, which brings the
fractions of the polymer chain further apart with time.
\begin{figure}[ht]
\vskip 1.0cm
\begin{center}
\includegraphics[scale=0.37]{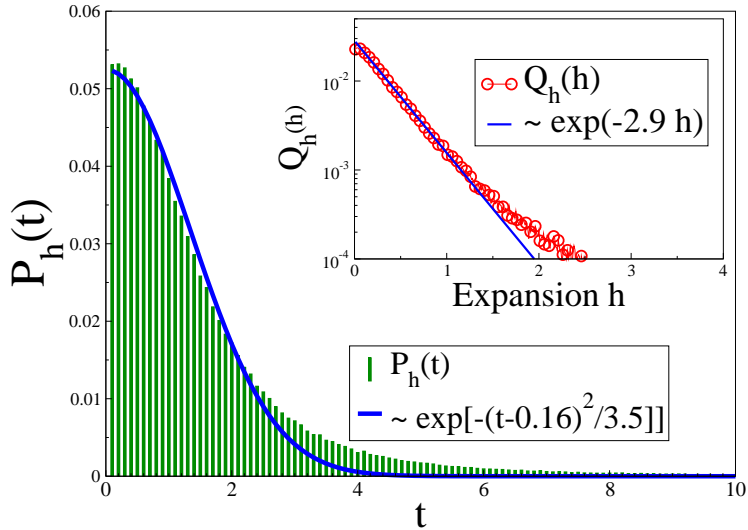}
\vspace{0.3cm}
\caption{Probability distribution $P_h(t)$ of  maximal times $t$ (impulses), and
$Q_h(h)$ - of maximal bond lengths $h$ (circles, cf. inset) before a
recombination  / healing event in a chain with $N=30$, $T=0.1$, $\gamma=0.25$
takes place. $P_h(t)$ is fitted by a Gaussian function, centered around a
{\em non-zero} value, $t = 0.16$, blue (thick) line; and $Q_h(h)$ indicates
exponential decay - blue (thin) line. \label{healing_fig} }
\end{center}

\end{figure}

In Fig.~\ref{healing_fig} we show the probability distribution of healed
bonds $P_h(t)$ against time $t$ elapsed after the bond length has exceeded the
threshold distance $r_h = 5.0$. Evidently, $P_h(t)$  is well approximated by a
Gaussian distribution except for the longer times $2 \le t \le 6$.
Interestingly, the maximum of the Gaussian is centered at a positive value of $t
\neq 0$, meaning that recombination begins with some time lag after the bond
length has gone beyond $r_h$ due to inertia. A related distribution of the bond
extensions $Q_h(h)$ beyond the one that corresponds to the maximal restoring
force $f^{max}$ (that is, beyond the inflection point of the Morse potential) is
shown in the inset of Fig.~\ref{healing_fig}. It is seen that the chance for
self-healing decreases exponentially fast with the stretching $h$ whereby a
deviation from the exponential relationship can be detected only for $Q_h  <
10^{-3}$. It becomes thus clear that for the adopted threshold of $r_h = 5.0$
the probability for subsequent recombination of a broken bond is vanishingly
small.

Generally, on the ground of the afore-mentioned results it appears that one
deals with a thermally activated breakdown  of a single macromolecule,
manifested by random uncorrelated events of bond scission. These results provide
no clue, however, to what extent non-linear effects due to anharmonic
interactions may play a role in thermolysis by inducing some degree of
collectivity in rupture events.

\subsection{Time-evolution maps of bond scission}

We tried to shed more light on the process of chain scission by producing
time-evolution maps which display the variation of different physical
quantities, characteristic for all individual bonds in a polymer chain, as a
function of time.  The idea behind this approach is to detect possible
collective excitations which might be responsible for bond breakage. It is
believed that a localized collective excitation will manifest itself by moving
with time along the chain so that a path on the time map would indicate its
location.

\begin{figure}[ht]
\includegraphics[scale=0.35, angle=180]{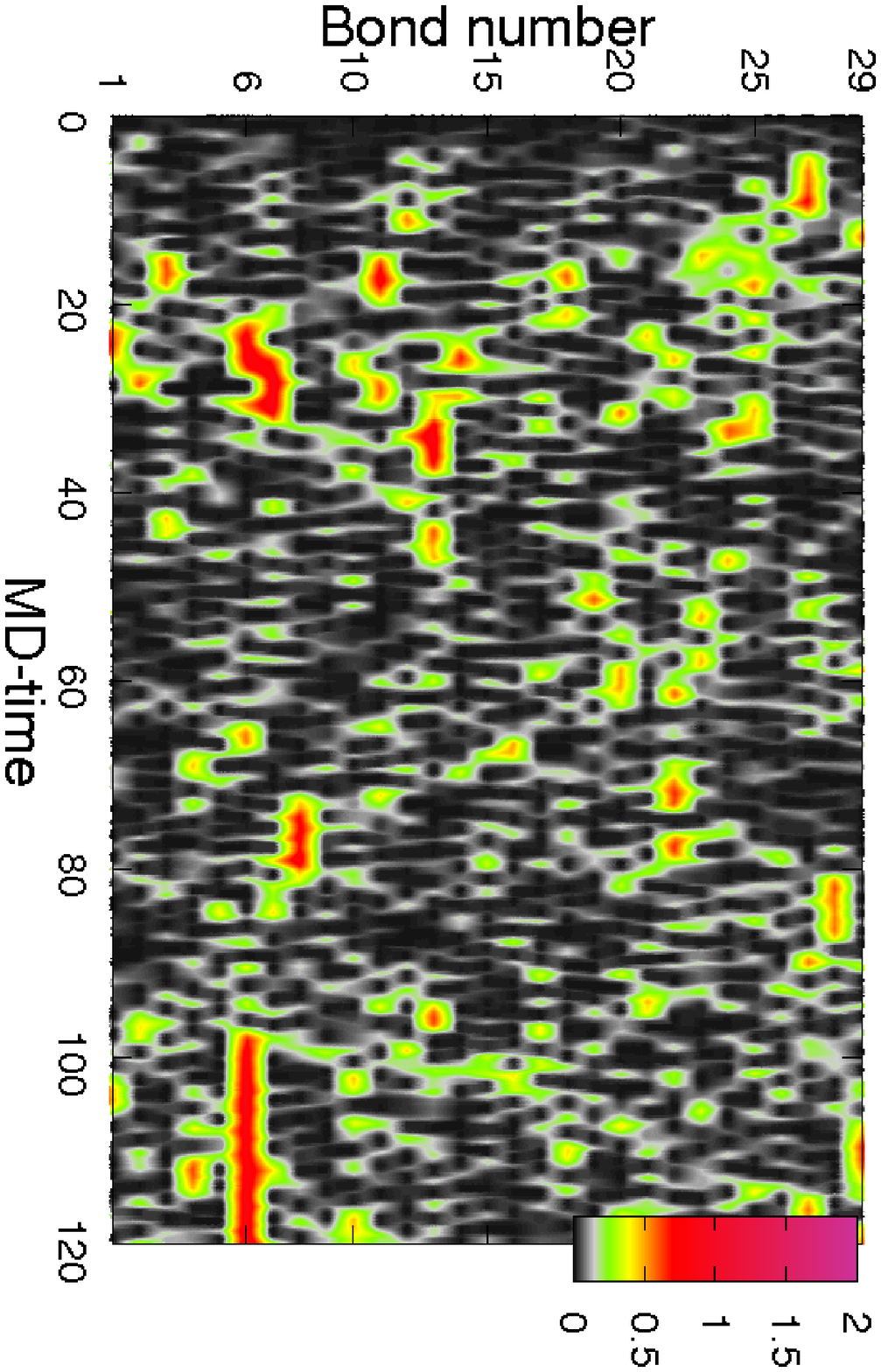} \includegraphics[scale=0.35,
angle=180]{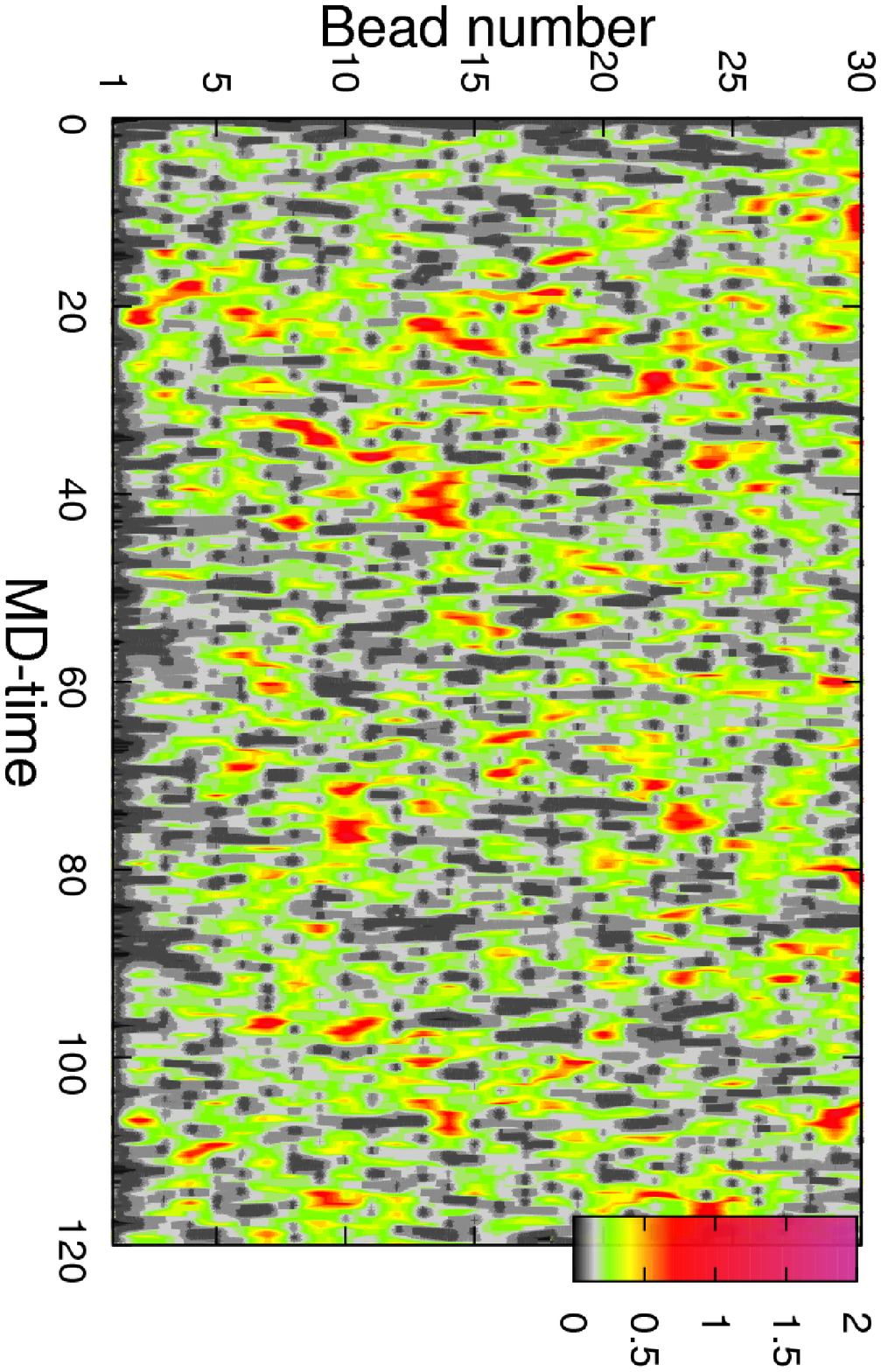} \caption{Color maps showing the time evolution of the
potential energy (left panel), and the kinetic energy (right panel) of
successive bonds / beads in a $N=30$ chain that breaks at time $t \approx 100$\;
t.u. for $T=0.1, \gamma = 0.25$.  Different colors indicate the value of the
energy as indicated in the legend. \label{poten_map_fig}
\vspace{0.2cm}}
\end{figure}

In Fig.~\ref{poten_map_fig} (left) we show the potential energy of successive
bonds along the polymer backbone  as a function of elapsed time after the start
of the simulation. The number of repeatable units is $N=30$ and the first
segment is tethered to a fixed position during the simulation while the $30$-th
segments moves freely.  The variation of the kinetic energy of individual beads
with time is shown in Fig.~\ref{poten_map_fig} (right).

A visual inspection of both maps suggests the lack of any coherence between the
local values of the kinetic and potential energies!  On the potential energy
map (which indicates the distribution and time evolution of strain along the
polymer backbone) one can clearly detect several events of bond scission.
Evidently, several bonds break and recombine in the time interval $10 \div 80\;
t.u.$ A real bond rupture, however, takes place in the $6$-th bond at $t \approx
100\; t.u.$ and the chain remains broken until the end of the observation time.
Notwithstanding, it is hard to detect a well expressed pattern in the
potential energies of the individual bonds that would indicate the existence
localized non-linear excitation (a soliton) as a possible precursor to a
scission event.
It appears that close to the tethered end of the chain, bond $1$, the potential
energy of the bonds is generally lower than in the other half of the chain. From
the map of the kinetic energy, however,  it is impossible to detect  time and
position of the breaking event at all. One merely sees that the fixed end of the
chain has almost zero kinetic energy, in contrast to the free end whose kinetic
energy remains high. Quiescent segments (dark spots) occur mainly  inside the
chain. As mentioned before, the map of the bond strain resembles strongly that
of the potential energy and is, therefore, not displayed here.

\begin{figure}[ht]
\includegraphics[width = 0.34\textwidth]{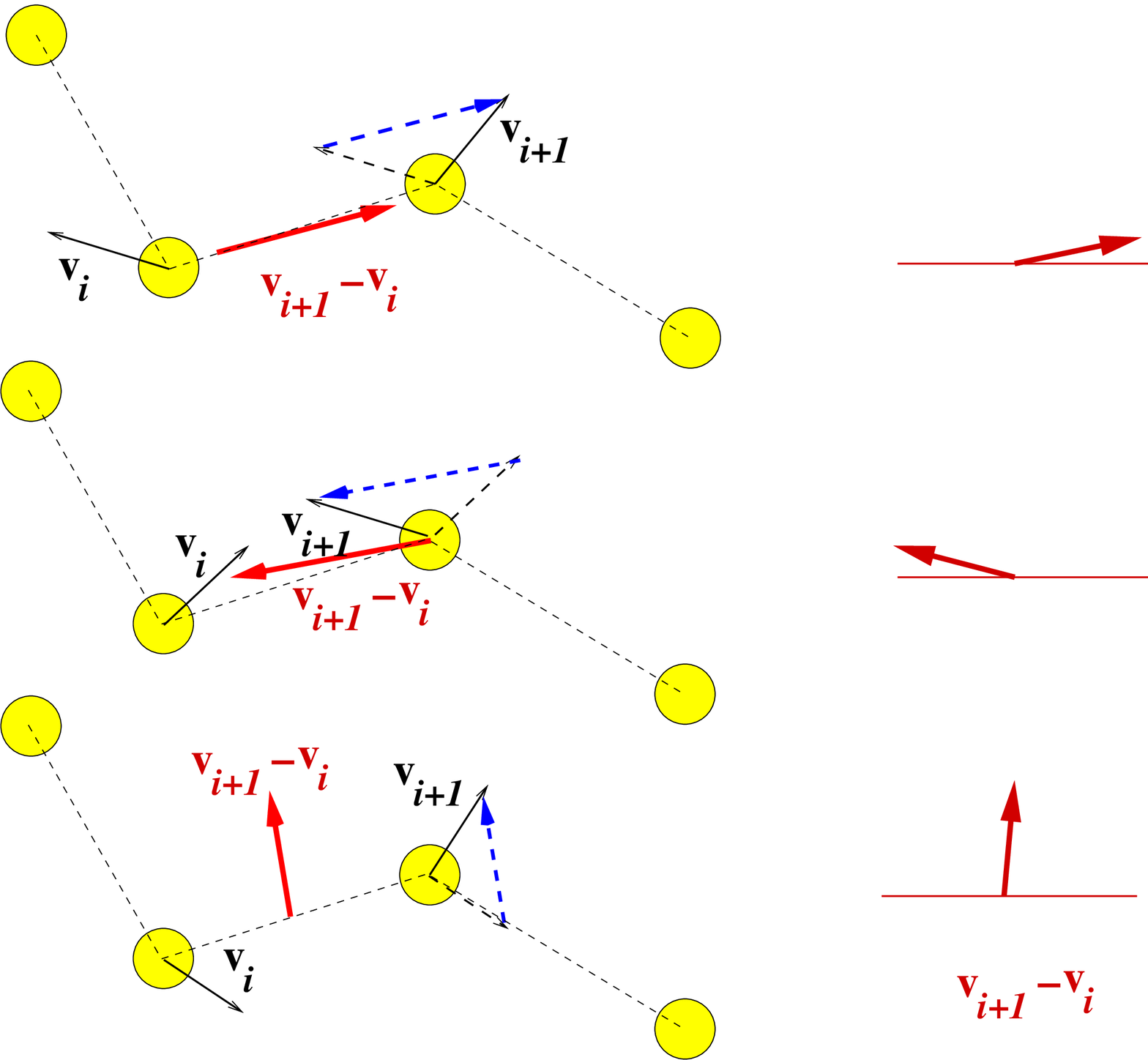}\hspace{1.5cm}
\vspace{1.0cm}
\includegraphics[scale=0.4,totalheight = 0.50\textwidth, origin =
br, angle=270]{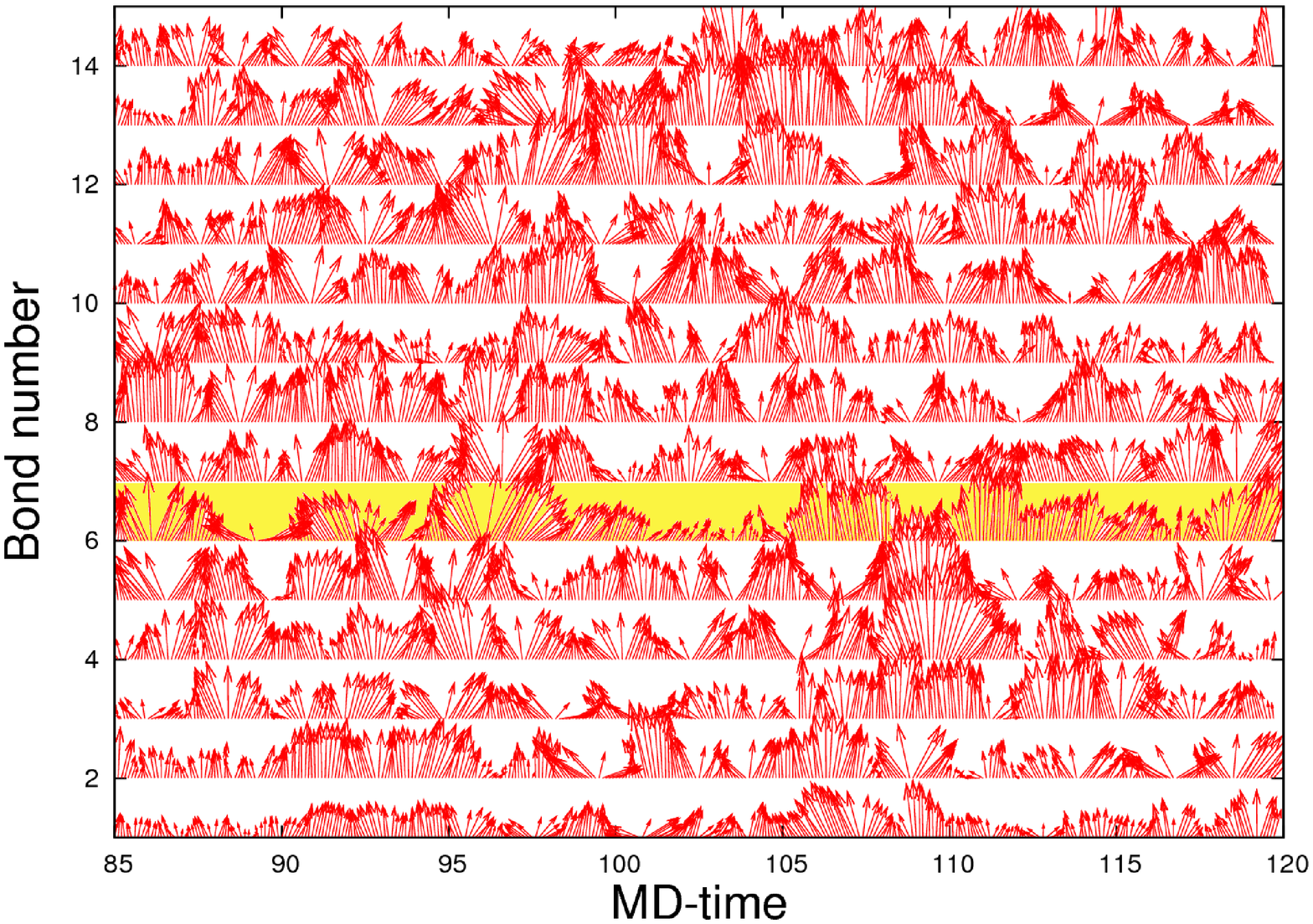}
\caption{(left panel) A sketch of the relative velocity $\Delta \vec{v}_i =
\vec{v}_{ i+1} - \vec{v}_{ i}$ red (full) line, of two beads that make a bond,
with respect to the bond vector, $\Delta \vec{r}_i =  \vec{r}_{ i+1} - \vec{r}_{
i}$.  (Anti) parallel orientations indicate expansion/shrinking  of the bond
whereas  $\Delta \vec{v}_i  \perp \Delta \vec{r}_i $ indicates bond rotation.
(right panel) Time evolution of bond stretching / contraction velocities,
$\Delta \vec{v}_i $ for all bonds of a polymer chain of length $N=30$ over time
interval $85 \le t \le 120$. The breaking bond, $n = 6$, marked with the yellow
color. Parameters are as in Fig.~\ref{poten_map_fig}. \label{arrow_fig}}
\end{figure}

An even more detailed picture of the individual motion of the monomers during
thermal destruction is provided by Fig.~\ref{arrow_fig} where in the map we
plot the {\em relative} velocities of the segments, connected by a particular
bond, over a time $85 \le t \le 120$. The map shows the first fifteen bonds
where a scission takes place. As explained in the scheme in
Fig.~\ref{arrow_fig}, the vector of the
relative velocity $\Delta \vec{v}_i = \vec{v}_{i+1} - \vec{v}_{ i}$  is oriented
perpendicular to the bond vector $\Delta \vec{r}_i =  \vec{r}_{ i+1} - \vec{r}_{
i}$  when such a bond rotates. Parallel (anti-parallel) orientation of  $\Delta
\vec{v}_i $ with respect to $\Delta \vec{r}_i $  indicates shrinking or
stretching of the bond $i$. As is evident from the velocity map, the latter
occur seldom and with small relative velocity. Predominantly the bonds are seen
to perform rather fast rotations regarding their previous orientation. Again,
the breaking bond (whose evolution is shaded yellow in Fig.~\ref{arrow_fig})
shows no distinct behavior with regard to the other bonds in the chain.
Altogether, no particular pattern of collective excitations can be detected
from the velocity map - nothing indicates the imminent failure of endangered
bond.

\subsection{Preferential scission of bonds along the backbone}

It turns out that considerably more insight into the nature of thermally
activated chain rupture may be derived from the probability distribution of
scission events along the backbone of the chain.

\begin{figure}[ht]
\begin{center}
\includegraphics[scale=0.34]{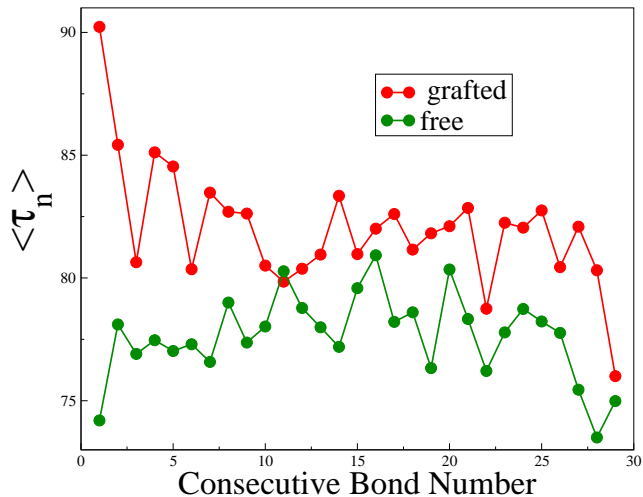}
\vspace{0.3cm}
\caption{Mean first breakage time $\langle \tau_n \rangle$ vs consecutive
bond number for a chain with $N=30$, $T=0.1$,
$\gamma=0.25$.\label{tau_bond_fig}}
\end{center}
\end{figure}

In Fig.~\ref{tau_bond_fig} we show the distribution of the MFBT $\langle \tau_n
\rangle$ of the individual bonds $n$ for a chain with $N=30$ in the under-damped
regime $\gamma = 0.25$ where inertial dynamics matter. We compare two cases: an
entirely free polymer chain and a tethered chain where one of the end-monomers,
$n = 1$, is fixed at a certain position in space. The simulation data is
characterized by considerable disarray as compared to cases with applied
external tensile force \cite{Ghosh} despite the very large statistics (more that
a total  of $1.5\times 10^6$ scission events) that was achieved in the computer
experiment.
For the free chain we find that $\langle \tau_n \rangle$ of the bonds at both
chain ends is about $10\%$ shorter, {\em if} such a bond is going to break. As
we shall see below, however, terminal bonds seldom happen to break. As
expected, the distribution of $\langle \tau_n \rangle$ is symmetric with respect
to the middle of the chain. In the case of a tethered chain, the bond in the
neighborhood of grafting bead survives longer than those in the middle. The bond
at the free end of the  chain is the most short lived one.

\begin{figure}[ht]
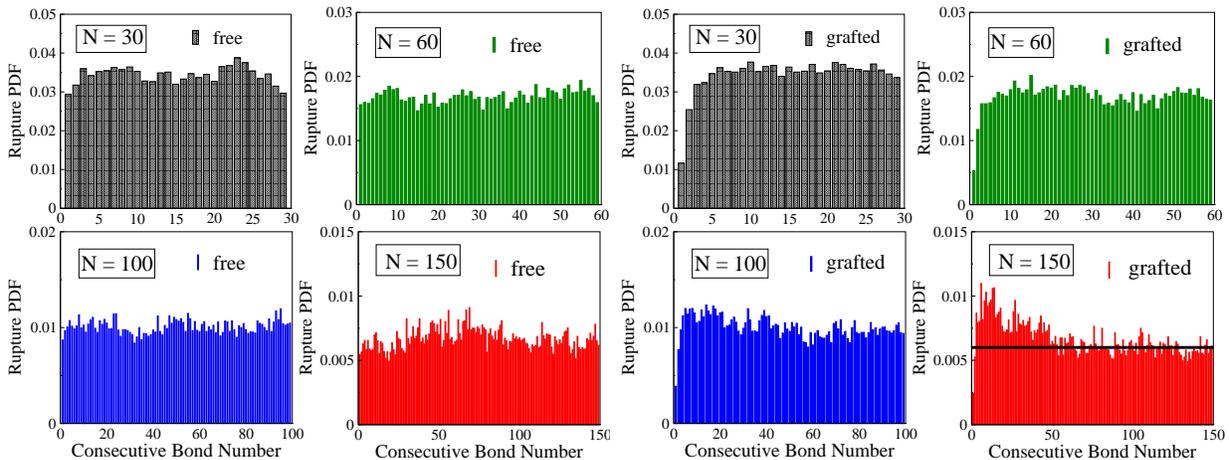

\includegraphics[scale=0.3]{rupture_histograms_free.eps}
\includegraphics[scale=0.3]{rupture_histograms_grafted.eps}
\caption{Overview of rupture probability vs consecutive bond number for a chains composed of
$N=20,40,60$ and $80$ beads ($T=0.1$, $\gamma = 0.25$).
\label{bond_histo_n_fig} }
\end{figure}

The non-uniformity in the distribution of scission events along the polymer
backbone is well pronounced also in the probability histograms,
Fig.~\ref{bond_histo_n_fig}, where
we display the (normalized) probability that a certain bond $n$ along the
polymer backbone will break within a time interval. Both grafted and free
chains are compared for different chain lengths $N = 30,\;60,\;100$ and $150$.
The asymmetry between scission rates for bonds that are respectively closer to
the fixed, or the free end of the polymer, is well pronounced. Two salient
features of the probability histograms, shown in Fig.~\ref{bond_histo_n_fig},
appear
characteristic:
\begin{enumerate}
 \item[i)] a well expressed  modulation of the rupture frequency along the
consecutive bond number, which is best visible in the case of a free chain with
$N=30$. There one observes a modulation within an interval that seems
to encompass roughly $15$ bonds. In longer chains such  modulation still
persist albeit the periodic pattern gets distorted.

\item[ii)]  For grafted chains a modulation also exists, however, one observes
a build-up of rupture probability  in the immediate vicinity of the fixed
monomer even
though the very first few bonds hardly break. This effect is most pronounced
in the longest chain $n = 150$.
\end{enumerate}

While the probability histograms, presented in Fig.~\ref{bond_histo_n_fig},
unambiguously indicate the existence of persistent differences in the likelihood
of bond breaking in regard with the consecutive number of a particular bond
along the chain backbone, the origin and the physical background of such
inhomogeneity is not self-evident. On the ground of the observed modulation of
the scission probability  one may speculate  that this self-induced
inhomogeneity results from the interplay of  thermal noise and the nonlinearity
of the bond-potential. It is furthermore conceivable that both control
parameters  such as the temperature $T$ and the nonlinearity of the
interactions, as well as other factors (e.g., the friction $\gamma$, i.e., under
/ over-damped dynamics) would affect the multi-scale-length inhomogeneity.
Presumably this finding presents an example of  more general phenomena where
different spatio-temporal order induced by the noise in nonlinear systems
\cite{Sagues}.

\begin{figure}[ht]
\begin{center}
\includegraphics[scale=0.3]{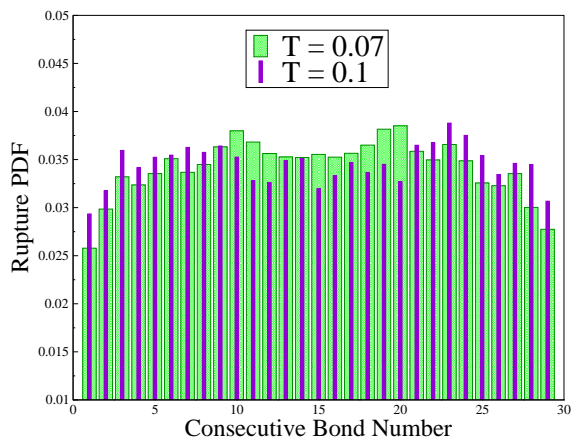}
\caption{Comparison of rupture histograms for grafted chains degraded at
different temperature of the heat bath. $N=30$, $\gamma =
0.25$.\label{bond_histo_T}}
\end{center}
\end{figure}

Since temperature is a major factor in thermal degradation, we present in
Fig.~\ref{bond_histo_T} the probability histogram for two different
temperatures, $T=0.07$ and $T=0.1$. One can see that the rupture histogram
visibly changes shape due to temperature. Evidently, a decrease of temperature
results in changing the positions of the local maxima which shift  closer to
each other while the modulation of rupture PDF grows.
Also, at lower temperature the histogram becomes less flat and the
non-uniformity in the rupture probability  increases.

Another interesting point concerns the different impact of under- (over)-damped
dynamics on the distribution of breaking events along the polymer backbone.
\begin{figure}[ht]
\begin{center}
\includegraphics[scale=0.3]{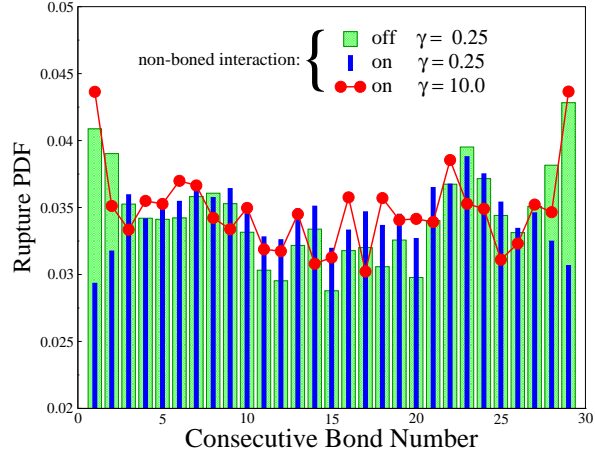}
\caption{Rupture histograms for free chains composed of $N=30$ particles
bonded by Morse potential in the under-damped, $\gamma = 0.25$, and overdamped,
$ \gamma = 10$, regimes. The presence / absence of {\em non-bonded}
interactions (indicated, respectively, as ``on'' and ``off'') has also a
dramatic effect on bond scission!  \label{bond_histo_nonbond}}
\end{center}
\end{figure}
The impact of friction $\gamma$  is displayed in Fig.~\ref{bond_histo_nonbond}
along  with the effect which  non-bonded interactions exert on the rupture
probability. Indeed, it is evident from Fig.~\ref{bond_histo_nonbond} that a
change of the system dynamics from under- to over-damped one by a
$40-$fold increase in $\gamma$ leads to qualitative changes in bond breakage.
The most striking effect of dynamics is the change in the frequency of scissions
at both ends of a free chain. When inertial effects are strong, terminal bonds
are the least likely to break while in the overdamped regime these become the
most vulnerable ones!  The well-pronounced dip in between the two maxima for
$\gamma=0.25$ also appears to vanish in the case of strong friction.

One should note that a  shape of the probability histograms, as for $\gamma =
0.25$, with the terminal bonds being the most resistant to rupture, has been
inferred from the experiments of Sheiko et al. \cite{Maty} on adsorption-induced
thermal degradation of carbon-carbon bonds on mica. The distribution of fragment
lengths with time in the course of thermolysis implies  a more or less constant
scission probability along most of the backbone while at the ends this
probability drops significantly \cite{Maty}. A similar conclusion is suggested
by the ultrasonic degradation experiments of Glynn et al. \cite{Glynn} with
polystyrene who found that the bonds in the middle of the chain break
preferentially to those at the ends. Probably, in the case of under-damped
dynamics over-stretched terminal bonds can quickly restore their equilibrium
length when friction is low and the restoring force needs only to pull back few
segments at the chain end.

In contrast, if chain motion is heavily damped, one might argue that the
over-stretched terminal bonds of a free chain need comparatively significantly
more time  to attain their normal length.  During this time they are longer
exposed to destructive thermal kicks which makes them more likely to break. This
would explain the increased vulnerability of bonds the closer they are  to the
chains ends in the over-damped dynamic regime.

Another interesting effect underlines the {\em role of the non-bonded neighbors}
in the rupture kinetics is also shown in Fig.~\ref{bond_histo_nonbond}. One can
readily verify from the respective histogram that the terminal bonds break
nearly $30\%$ more often as those in the middle of the chain when the non-bonded
interactions are switched off. In some sense, one could claim that the
non-bonded interactions, when absent, render the chain breakage similar to what
one has in the over-damped case.

In order to gain more insight into this behavior we recall that  in a good
solvent the mean squared end-to-end distance of a polymer chain $R_e^2$ is
significantly larger than the radius of gyration, $R_g^2$, that is, $R_e^2
\approx 6R_g^2$. This suggests that the terminal beads 'live' predominantly at
the outskirts of the polymer coil where the chance for collision with another
monomer is reduced. Excluded volume interactions are thus weaker on the average,
and, correspondingly, terminal bonds are on the average less stretched, i.e.,
they are less likely to break.

\begin{figure}[ht]
\begin{center}
\includegraphics[scale=0.3]{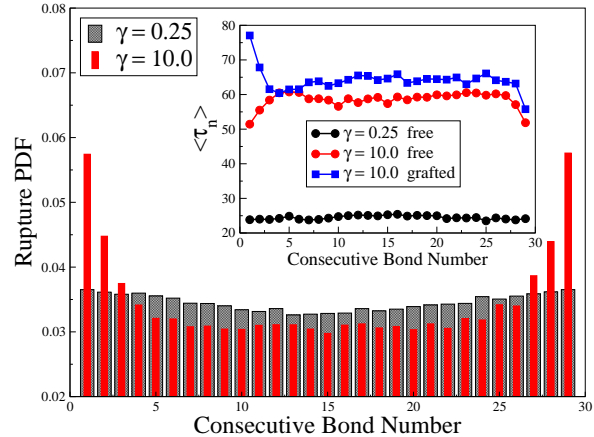}
\caption{Rupture histograms of a $30$-particles free harmonic chain for under-
($\gamma = 0.25$) and over-damped cases ($\gamma=10$). The inset shows mean
values of breakage times with respect to consecutive bond number for grafted and
free chains. ($T=1.5$)}
\label{harmonic}
\end{center}
\end{figure}

As far as our Langevin MD simulation deals essentially with anharmonic (Morse
and Lennard-Jones) interactions between chain monomers, it appears useful to
make a comparison with a reference system, a purely harmonic chain Fig.~
\ref{harmonic}, that has been studied recently \cite{Lee,Sokolov}.
As in \cite{Sokolov}, the threshold $r_h$ for rupture  of such a Gaussian chain
is set arbitrary to some extension of the harmonic bond -  bonds longer than
$r_h$  are then considered broken.  It is seen from Fig.~\ref{harmonic}, that
shape of the  rupture histogram in the overdamped case, $\gamma = 10$, is
similar to that of our Morse chain in the same dynamic regime, and resembles
closely the histogram shape in Fig.~\ref{bond_histo_nonbond} when non-bonded
interactions are switched off. Our simulation result reproduces very well the
recent observations of Fugmann and Sokolov \cite{Sokolov},  cf. Fig.~2 in
\cite{Sokolov}, who modeled the thermally induced breakdown of a Gaussian
chain.
\begin{figure}[ht]
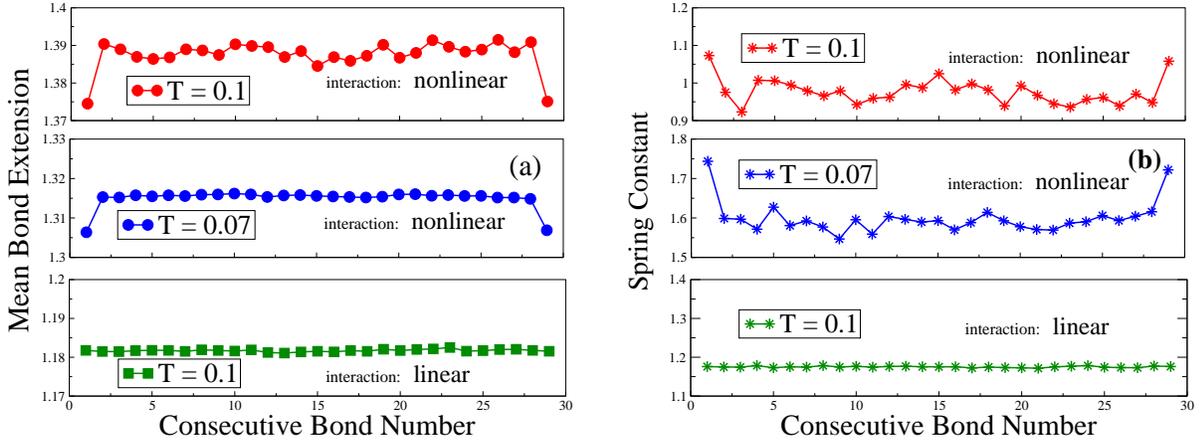

\begin{center}
\hspace{-1.0cm}
\includegraphics[scale=0.3]{b.eps}
\hspace{0.5cm}
\includegraphics[scale=0.3]{springs.eps}
\caption{Mean extension of bonds $\langle b_n\rangle$ (left panel) and mean
effective spring constant $\lambda_n =  k_BT /\langle \Delta b_n^2\rangle$ of
the individual bonds  (right panel) vs. consecutive bond number $n$ for a free
chain with $N=30$-particles. Here $T=0.1$, $\gamma = 0.25$. }\label{bond_exten}
\end{center}
\end{figure}
In the under-damped regime, however, which was not studied before, the Gaussian
chain histogram is strongly leveled off, retaining only a very weak (symmetric)
increase in the rupture probability  of the individual bonds as one moves away
from the
center of the chain and approaches the free ends. Any trace of self-induced
inhomogeneity as in the case of non-linear interactions,
Fig.~\ref{bond_histo_n_fig}, is absent. This supports again our assumption that
the observed inhomogeneity in the rupture probability  distribution among
individual
bonds occurs as a result from interplay between the thermal noise and the
nonlinearity.

A test of this conclusion is suggested by the study of another property - the
average strain of the bonds $\langle b\rangle$ with respect to the consecutive
bond number, shown in Fig.~\ref{bond_exten}. One can see that this quantity
resembles the behavior of rupture probability regarding $n$. The terminal bonds
are less stretched than the other ones and therefore break seldom. Moreover, the
the {\em effective} spring constant $\lambda_n$ of the individual bonds which is
given by the variance  of the strain, $\langle \Delta b_n^2\rangle$, shown in
the right panel in Fig.~\ref{bond_exten}, behaves similarly. Indeed, the
Hamiltonian $H = \frac{1}{2} \sum_{n=1}^{N-1} \lambda_n (\Delta b_n)^2$ of the
chain defines a  bond length probability distribution function ${\cal P}$
$\propto \exp\{-H / k_BT\}$. The distribution of individual strain is Gaussian,
${\cal P}_n (\Delta b_n) \propto \exp[- (\Delta b_n)^2 / k_BT  \langle \Delta
b_n^2 \rangle ] $ with  $\langle \Delta b_n^2\rangle = k_BT / \lambda_n$. Thus
one can see that in an uniform chain there appear regions of effectively
``stiffer'' bonds (at the ends of the chain), and of ``softer`` bonds (away from
the ends) that are less or more likely to stretch and break, respectively.
Remarkably, in a harmonic Gaussian chain, where excluded volume effects are
absent, both $\langle b\rangle$ and $\lambda_n$ are seen to be entirely
uniformly distributed, Fig.~\ref{bond_exten} (lowest panels)! This proves
that the observed inhomogeneity is indeed attributed to the nonlinearity of the
bond potential.

\subsection{Molecular Weight Distribution during Thermal Degradation}

\begin{figure}[ht]
\begin{center}
\hspace{-1.0cm}
\includegraphics[scale=0.67]{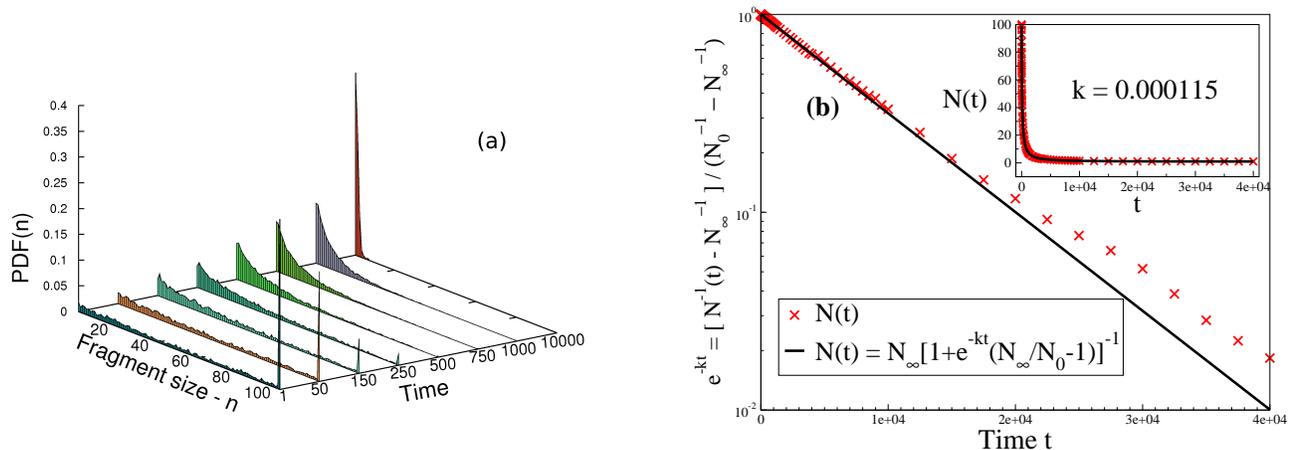}
\hspace{0.5cm}
\includegraphics[scale=0.3]{Nfrag.eps}
\caption{(a) Probability distribution of fragment sizes $P(n,t)$ at different
times $t$ (in MD time units) after beginning of the thermal degradation process
for a chain of length $N(t=0) = N_0 = 100$. At late times $t \approx 10^4\;
t.u.$
the chain disintegrates into single segments, $N_\infty = 1$. (b) Variation of
the mean fragments length $N(t)$ after the onset of thermal degradation. The
solid line denotes the theoretical result, Eq.~(\ref{decay}) with respect to
$\exp(-k t)$ and proves that the process evolves predominantly as a first order
chemical reaction. The inset shows $N(t)$ vs $t$ in normal coordinates. One
finds here $k = 1.15\times 10^4$ [t.u.$^{-1}$]. }\label{fragment}
\end{center}
\end{figure}

Finally, we examine the course of the degradation kinetics which is manifested
by a time-dependent probability distribution function, $P(n,t)$, of the
fragments of the initial macromolecule as time elapses after the onset of the
process - Fig.~\ref{fragment}. The initial length of our polymer is $N=100$ and
the temperature $T = 0.10$. Averages of $P(n,t)$, obtained over $10^4$ cycles
of scission - Fig.~\ref{fragment}a - are shown to evolve in time from a
$\delta-$like distribution at $t=0$ to a rather flat distribution with a rapidly
growing second maximum around sizes of $n \approx 2\div5$. After a short time of
$\approx 250$ t.u., the initial chain has  already disintegrated into small
clusters whose length is most probably $n \approx 1$.  At late times, $t \approx
10^4\; t.u.$, the distribution $P(n,t)$ attains again a sharply-peaked
$\delta-$like shape. We would like to stress that the observed variation of
$P(n,t)$ resembles strongly the one, found in recent experiments with bond
scission in poly(2-hydroxyethyl methacrylate) chains \cite{Maty,Lebed}. If one
assumes that the scission kinetics is described by a first-order reaction, then
one may derive an analytic expression for the decrease in the average length of
the fragments with elapsed time \cite{Lebed} as
\begin{equation} \label{decay}
 \left ( \frac{1}{N(t)} - \frac{1}{N_{\infty}} \right ) =  \left (
\frac{1}{N_0} - \frac{1}{N_{\infty}} \right ) e^{-k t},
\end{equation}
where $N_0$ is the initial contour length at $t=0$ and $N_{\infty}$ is the mean
contour length of polymer chains at infinite time. $k$ is a first-order rate
constant. This result, Eq.~(\ref{decay}), is compared with our simulation
data in Fig.~\ref{fragment}b. In normal coordinates (inset) the course of the
theoretical result can hardly be distinguished from from the simulation data.
The final fragment size is $N_\infty = 1$. Upon a closer look in semi-log
coordinates, however, the exponential decay $\propto \exp(-k t)$ is found to
deviate slightly at late times $t > 2\times 10^4$\; t.u. We interpret the
observed
discrepancy as an indication that fragment recombination may occasionally take
place at late times when the fragments become sufficiently small and mobile.
Recombination comprises a second-order (binary) reaction which adds to the
dominant first-order reaction of decay and, therefore, contributes to the
observed deviations. Nonetheless, it appears that our simulation model
faithfully accounts for the degradation kinetics.

\section{Concluding remarks} \label{Conclusions}

In this work we presented a MD simulation study of the thermal degradation of a
single polymer molecule at different conditions, such as temperature, friction,
length of the polymer, etc. A major emphasis has been placed on exploring the
effect of  the nature (linear or not) of the monomer interactions on the
scission process.

According to our results, the basic notion of the fragmentation process as a
result of random and independent scission of bonds  is supported by the
observed dependence of the MFBT $\langle \tau \rangle$ on the chain length
$N$ as  $\langle \tau \rangle \propto 1 / N$.  In addition, the variation of
the life time, $\langle \tau \rangle$, with (inverse) temperature turns out to
be an Arrhenian-law, $\langle \tau \rangle \propto \exp (\Delta E_b/k_B T)$
whereby the activation energy is rather close to the potential well depth of the
Morse interaction.

A very detailed investigation of the  time-evolution maps of bond and segment
properties during thermolysis does not indicate any  existence of  some kind of
localized nonlinear excitations (solitons) which would mediate the energy
transfer and cause bond breakage.

In contrast, the distribution of the bond rupture probability reveals the
existence of a multiple length-scale inhomogeneity which is self-induced
presumably as a result of the interplay  between thermal noise and the
nonlinearity of the bond potential. This inhomogeneity does not exist in
 the Gaussian chain model where the forces depend linearly on distance
between monomers which supports the notion of force anharmonicity (i.e.,
non-linearity) as an origin of the observed inhomogeneity.

In addition, two major factors have been found to effect the scission kinetics
of  a single polymer chain - (i) the presence and relative strength of excluded
volume   (i.e., non-bonded) interactions between monomers, and (ii) the
particular kind of chain dynamics (over- or under-damped regimes). Both effects
have been demonstrated to discriminate strongly between terminal bonds  and
bonds away from the chain ends as far as the probability of breakage is
concerned.

In this study we examine the fragmetation kinetics during thermal degradation
and obtain the time evolution of the fragment size distributionm after the
onset of the process. We find that the size distribution agrees well with
experimental observations. The time variation of the mean fragment size
indicates that the scission process can be well described by means of a
first-order reaction kinetics  with only small deviations due to recombination
events.

It is a matter of a future investigation to elucidate fully the nature of this
interesting phenomenon.

\section{Acknowledgments}

 We thank Peter Talkner for stimulating discussion during his visit in Mainz.
A.~M. gratefully acknowledges support by the Max-Planck-Institute for Polymer
Research during the time of this investigation. This investigation has been
supported by the Deutsche Forschungsgemeinschaft (DFG),
Grants SFB $625$/B$4$ and FOR $597$.


\begin{thebibliography}{99}
\bibitem{Allen} N. S. Allen and M. Edge, {\em Fundamentals of Polymer
Degradation and Stabilization}, Elsevier Applied Science, New York, 1966.
\bibitem{Madras} G. Madras, J. M. Smith, and B. J. McCoy, Ind. Eng. Chem. Res.
{\bf 35}, 1795 (1996).
\bibitem{Nyden} M. R. Nyden, G. P. Forney, and G. E. Brown, Macromolecules,
{\bf 25}, 1658 (1992).
\bibitem{Flynn} J. H. Flynn and R. E. Florin, in {\em Pyrolysis and GC in
Polymer Analysis}, Eds. S. A. Liebman and E. J. Levy, Marcel Dekker Inc.: New
York, 1985, p. 149.
\bibitem{Schulten} M. Sarikaya, C. Tamerler, A. K. Jen, K. Schulten, and F.
Banyex, Nature Mater. {\bf 2}, 577 (2003)
\bibitem{Han} T. H. Han, J. Kim, J. S. Park, C. B. Park, H. Ihee, and S. O.
Kim, Adv. Mater. {\bf 19}, 3924 (2007)
\bibitem{Simha} R. Simha, J. Appl. Phys. {\bf 12}, 569 (1941)
\bibitem{Jellinek} H. H. G. Jellinek, Trans. Faraday Soc. {\bf 1944}, 266 (1944)
\bibitem{Ballauff} M. Ballauff and B. A. Wolf, Macromolecules, {\bf 14}, 654
(1981).
\bibitem{Ziff} R. M. Ziff and E. D. McGrady, Macromolecules {\bf 19}, 2513
(1986);  E. D. McGrady and R. M. Ziff, Phys. Rev. Lett. {\bf 58}, 892 (1987)
\bibitem{Cheng} Z. Cheng and S. Redner, Phys. Rev. Lett. {\bf 60}, 2450 (1988)
\bibitem{Nyden2} E. Blaisten-Barojas and M. R. Nyden, Chem. Phys. Lett. {\bf
171}, 499 (1990); M. R. Nyden and D. W. Noid, J. Phys. Chem. {\bf 95}, 940
(1991)
\bibitem{Doerr} T. P.  Doerr and P. L. Taylor, J. Chem. Phys. {\bf 101}, 10107
(1994)
\bibitem{Wang} M. Wang, J. M. Smith, and B. J. McCoy, AIChE Journal, {\bf 41},
1521 (1995)
\bibitem{Hathorn} B. C. Hathorn, B. G. Sumpter, and D. W. Noid, Macromol.
Theory Simul. {\bf 10}, 587 (2001)
\bibitem{Doruker} P. Doruker, Y. Wang, and W. L. Mattice, Comp. Theor. Polym.
Sci. {\bf 11}, 155 (2001)
\bibitem{Lee} C. F. Lee, Phys. Rev. E {\bf 80}, 031134 (2009)
\bibitem{Sokolov} S. Fugmann and I. M. Sokolov, Phys. Rev. E {\bf 81}, 031804
(2010).
\bibitem{Goldenfeld} N. Goldenfeld, {\it Lectures on Phase Transitions and the Renormalization Group},
Addison-Wesley Publishing Company, New York, 1992.
\bibitem{Gardiner} C.W. Gardiner, {\it Handbook of Stochastic Methods}, Springer-Verlag,
Berlin Heidelberg, 2004.
\bibitem{Ghosh} A. Ghosh, D. I. Dimitrov, V. G. Rostiashvili, A. Milchev and T.
A. Vilgis, J. Chem. Phys. {\bf 132}, 204902 (2010)
\bibitem{Sagues} F. Sagu\'es, J. M. Sancho, J. Garcia-Ojalvo, Rev. Mod. Phys. {\bf 79}, 829 (2007).
\bibitem{Maty} S. S. Sheiko, F. C. Sun, A. Randall, D. Shirvanyants, M.
Rubinstein, H.-I. Lee, and K. Matyjaszewski, Nature Lett., {\bf 440}, 191
(2006).
\bibitem{Glynn} P. A. R. Glynn, B. M. E. van der Hoff, and P. M. Reilly, J.
Macromol. Sci. - Chem. {\bf A6}, 1653 (1972)
\bibitem{Lebed}  N. V. Lebedeva, F. C. Sun,  H.-I. Lee, K. Matyjaszewski, and
S. S. Sheiko,  J. Am. Chem. Soc. {\bf 130}, 4228 ( 2007) ;  I.Park, S. S.
Sheiko,  A. Nese,  K.  Matyjaszewski, Macromolecules, {\bf 42}, 1805 (2009).
\end{thebibliography}
\end{document}